\documentclass[aip,graphicx]{revtex4-1}

\draft 
\usepackage{graphicx}
\begin{document}


\title{Quantum effects in the brain: A review} 



\author{Betony Adams}
 \affiliation{Quantum Research Group, School of Chemistry and Physics, and National Institute for Theoretical Physics, University of KwaZulu-Natal, Durban, 4001, South Africa.}
\author{Francesco Petruccione}%
\affiliation{Quantum Research Group, School of Chemistry and Physics, and National Institute for Theoretical Physics, University of KwaZulu-Natal, Durban, 4001, South Africa.}


\date{\today}

\begin{abstract}
In the mid-1990s it was proposed that quantum effects in proteins known as microtubules play a role in the nature of consciousness. The theory was largely dismissed due to the fact that quantum effects were thought unlikely to occur in biological systems, which are warm and wet and subject to decoherence. However, the development of quantum biology now suggests otherwise. Quantum effects have been implicated in photosynthesis, a process fundamental to life on earth. They are also possibly at play in other biological processes such as avian migration and olfaction. The microtubule mechanism of quantum consciousness has been joined by other theories of quantum cognition. It has been proposed that general anaesthetic, which switches off consciousness, does this through quantum means, measured by changes in electron spin. The tunnelling hypothesis developed in the context of olfaction has been applied to the action of neurotransmitters. A recent theory outlines how quantum entanglement between phosphorus nuclei might influence the firing of neurons. These, and other theories, have contributed to a growing field of research that investigates whether quantum effects might contribute to neural processing. This review aims to investigate the current state of this research and how fully the theory is supported by convincing experimental evidence. It also aims to clarify the biological sites of these proposed quantum effects and how progress made in the wider field of quantum biology might be relevant to the specific case of the brain. 
\end{abstract}


\maketitle 

\section{\label{sec:level1}Introduction}

The idea that quantum physics might have something to do with explaining that most mysterious organ, the brain, has generated scepticism among scientists. Just because quantum theory and consciousness are both complex concepts does not mean they necessarily inform each other. Indeed the contexts in which both occur have long been assumed to be incompatible. The brain is a biological system, functioning at physiological temperatures and subject to the myriad interactions by which living organisms survive. Quantum effects, on the other hand, are conventionally limited to low temperature systems isolated from the detrimental effects of environmental interaction. However, the mutual exclusion of biological and quantum systems is no longer so absolute. Research in the field of quantum biology has had some success in identifying how quantum processes might benefit living organisms \cite{alkhalili,mcfadden,mohseni,lambert,futureQB}. On a fundamental level, it can be said that all biological systems are quantum mechanical, being composed of atoms and thus subject to the quantum theory of atomic structure first developed by Bohr and Rutherford at the beginning of the twentieth century \cite{zettili,haken}. In the field of quantum biology these are considered to be trivial quantum effects, what is more interesting is whether quantum phenomena such as coherence, entanglement and tunnelling might play a non-trivial role in enhancing the efficacy of biological processes. There is mounting evidence that quantum coherence might contribute to the extraordinary efficiency of photosynthesis. The avian compass has also been suggested to exploit quantum effects with behavioural evidence supporting the hypothesis. Olfaction, enzyme catalysis and the intricacies of DNA have all fallen under the scrutiny of researchers working in the field of quantum biology \cite{alkhalili,mcfadden,mohseni,lambert,futureQB}. The basic biology of the brain, elevated though it is by the inexplicable phenomenon of consciousness, is perhaps not, on a mechanistic level, so very different from other processes that take place in the body. It is no wonder then that interest has grown in the detailed physiological mechanisms that constitute the central nervous system. Being that it is still not perfectly understood exactly how the brain and its related systems work, quantum theory might have at least some contribution to make. There have been attempts to tackle the hard problem of consciousness \cite{penrose,hameroff2014}, or apply quantum theory to human psychology and cognition \cite{jedlicka, bruza,busemeyer}. This review, however, focuses on the various instances in which it has been suspected that quantum effects play a role in the structural mechanisms by which the brain performs its integral functions: the firing of nerves; the actions of anaesthesia, neurotransmitters and other drugs; the sensory interpretation and organised signalling that is central to the vast neural network that we identify as our \emph{self}.
\subsection{\label{sec:level2}The classical brain}
Neuroscience has made great strides in understanding how the brain works. However, if our model of the brain were already fully realised there would be no need to investigate whether quantum theory might offer any insights. The question of how the brain works is, to state the very obvious, complicated. On one level it is a matter of \emph{matter}, the network of cells and signalling processes that constitute the central nervous and related systems. But there is also the question of how this physiology gives rise to the phenomenon of consciousness. While advances in imaging techniques have resolved some of the structure of the brain, new discoveries are still being made. Recently, previously unheard of lymphatic vessels were discovered in the meninges of the brain, suggesting a link between the central nervous system and the immune system and prompting research into their role in neurodegeneration \cite{louveau1,louveau2}. Even less well understood is the relationship between structure and function and to what extent functional connectivity, or an understanding of the mind, can be explained by the basic anatomy of the brain \cite{lynn,ramo,rodriguez,messe}. 

\subsubsection{Brain organisation}
The central nervous system is made up of the brain and the spinal cord. The human brain is a complex organisation of neural tissue, with approximately 86 billion neurons \cite{azevedo}. While neurons are responsible for the electrical activity of the brain they are supported by glial cells, which have a number of functions \cite{araque}. Brain matter consists of both grey and white matter, where the former is mainly cell bodies and the latter is predominantly made up of myelinated axons that allow for the transport of signals and the connectivity of the different brain sections \cite{tameem}. The brain is a network, dependent on the complex interaction of its different constituent parts, thus the assignment of specific function to specific region is a simplification. Nevertheless, for classification purposes it is often divided and subdivided into a number of regions, the simplest of these divisions being the fore, mid and hindbrain.

\begin{figure}
	\begin{center}
		\includegraphics{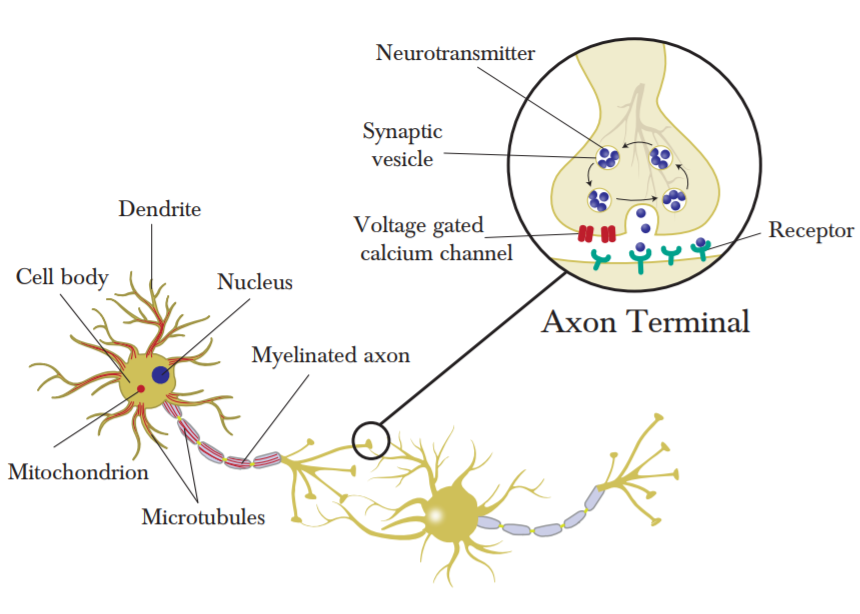} 
		\caption{The basic structure of a nerve cell. Although a number of the organelles are missing this simplified schematic is sufficient to gain some understanding of the location of the various quantum effects discussed in this review. Of interest are the microtubules which occur in different arrangements throughout the nerve cell and the mitochondria, which will be discussed in the context of electron transfer. The myelin insulated axon has also been proposed as a wave guide for neural photonics. The meeting of dendritic spine and axon terminal at the synaptic cleft is enlarged for clearer understanding of the mechanism of neurotransmission and the transfer of neural signals. Synaptic vesicles are also the site of the proposed uptake of Posner molecules during endocytosis.}\label{fig1_schematic}
	\end{center}
\end{figure}
\subsubsection{Neural action}
Nerve cells, the main constituents of the central nervous system, are elongated cells consisting of cell body, dendrites and axon. A complex network of neurons throughout the body allows for the propagation of information that is facilitated by the firing or not firing of neurons, the measure of a neuron's action potential. The biophysical mechanism of action potential in neurons is conventionally understood through the work of Hodgkin and Huxley \cite{hodgkin}. In order for a neuron to fire the resting potential must be raised to the requisite threshold potential. This is mediated by the gradient of electrically charged ions distributed across the cell membrane. Communication between neurons is integral to their ability to convey information. There is still some contention as to exactly how nerves communicate, with evidence for both chemical and electrical signal transmission \cite{pereda}. This review will focus on the former, where neural communication is mediated by the release of neurotransmitters. When an action potential propagating along a neuron reaches the axon terminal it triggers the opening of voltage gated ion channels which in turn stimulate exocytosis, or the release of neurotransmitters into the synaptic cleft \cite{lodish}. These neurotransmitters diffuse across the synaptic cleft and bind to special receptors on the dendritic spines, opening other ion channels. Ions can now enter this nerve cell and change the membrane potential, potentially generating an action potential in the post-synaptic nerve cell \cite{lodish}. While this is a simplistic model of the action of neurons it serves as a means to locate the quantum effects discussed in the next section. Although neurons have a number of constituent parts, those important in the context of this review are as follows. Microtubules, as the location of effects described in theories of consciousness as well as coherent quantum transport. Mitochondria, as the site of electron transport processes. The axon, as mediator of electrical signals and possible site of biophoton transfer. A basic understanding of the synaptic mechanism will also be beneficial to the discussion of both quantum effects in neurotransmission as well as the neural action of Posner molecules.  

\subsubsection{Consciousness}
While this review is preoccupied with the simpler question of structure, that is, the quantum physics of certain physiological mechanisms in the brain, it does include some theories as to how this physiology manifests as consciousness. Biology based attempts to explain consciousness include identifying its neural correlates by means of neuroimaging methods which study the changes in neural activity between conscious and unconscious states as well as altered states of consciousness \cite{cavanna,cavanna2,owen,rohaut,boly,aru}. A formal description of consciousness, given the difficulty of quantifying its subjective experience, would likely borrow from complex network theory as well as disciplines from physics and philosophy \cite{lynn,douglass,tononi,tononi2,xerxes,sousa}. The question is still open as to whether quantum physics has something to add to the debate.
\begin{figure}
	\includegraphics [scale = 0.5]{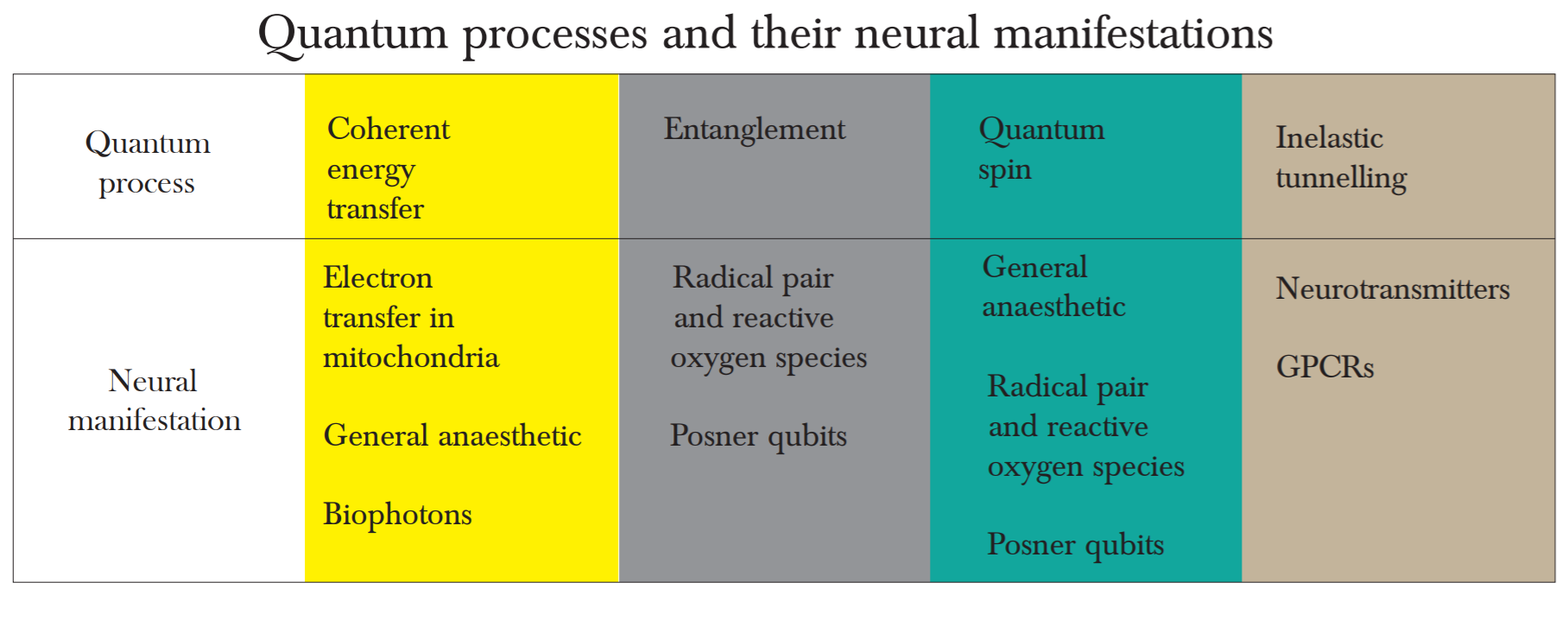}%
	\caption{The non trivial quantum effects discussed in this review and their neural context.\label{}}%
\end{figure}

\subsection{\label{sec:level3}Non trivial quantum effects}
In 1913, Bohr presented his model of the atom where, in contrast to classical orbits, electrons occupied discrete energy levels. This followed quickly on the heels of Planck's advances in understanding blackbody radiation and Einstein's explanation of the photoelectric effect which, along with Compton's work with X-rays, ushered in the new era of quantum mechanics \cite{zettili,haken}. Fundamentally, all biology can be described as being quantum mechanical in the same way that all matter is quantum mechanical. However, the aim of this review is to investigate non trivial quantum effects in biological systems, to widen the scope of quantum theory to include biological mechanisms and not merely the description of their constituent atoms. Quantum weirdness is a well documented phenomenon. First Planck and Einstein demonstrated that radiation, normally understood as behaving like a wave, can also behave like a particle. De Broglie then suggested that matter, which seems discrete, can sometimes show wave-like effects such as interference. Despite being famous for its uncertainty, the formalisation of the theory has proved extremely successful in describing the behaviour of microscopic systems. The mathematical framework of quantum mechanics associates a physical system with a quantum state that contains all possible information on the system. What is interesting is that within this framework, for two quantum states describing a system, a linear combination of these states also describes the system. It is this that gives rise to the uniquely quantum effect of superposition states, one of the non-trivial effects discussed in this review. The concept of quantum coherence, which quantifies the relationship between states in a superposition, is investigated in various biological contexts. As is entanglement, the non-classical correlation between different quantum states. Entanglement is often discussed with reference to quantum spin, the property of elementary particles that determines their behaviour in a magnetic field. Electron tunnelling is also raised as a potential candidate for quantum effects in biological systems. An explanation of tunnelling follows from the probabilistic description of quantum mechanics which allows the possibility of a quantum particle passing through a classically forbidden potential barrier \cite{zettili,haken}. This review is particularly focused on inelastic tunnelling, in which the tunnelling electrons are coupled to vibrational modes in their biological context.

\section{\label{sec:level1}Quantum effects in neural processes}

It is perhaps misleading to talk about quantum processes in the brain. Nerve cells extend throughout the body. The quantum processes discussed in this review are not confined to the brain but take place within neurons and at synapses. They are therefore implicated in the biological functioning of the entire body. It might be more accurate to describe these effects as quantum enhanced neural processing.
\subsection{\label{sec:level2} Orchestrated Objective Reduction}
Orchestrated objective reduction (Orch OR), the application of quantum mechanical formalism to the question of consciousness, was proposed by Stuart Hameroff and Roger Penrose in the 1990s \cite{hameroff1996a,hameroff1996b,hameroff2014}. In his 1989 book \emph{The Emperor's New Mind: Concerning Computers, Minds and The Laws of Physics} Penrose addresses the possibility that the laws of classical physics are not sufficient to explain the phenomenon of consciousness, suggesting instead that quantum physics might be integral to this explanation \cite{penrose}. His hypothesis initially lacked a biological context in which these quantum effects might occur. Hameroff, an anaesthesiologist by training, had been previously interested in microtubules and suggested that they could be a contender in which to situate a quantum model for consciousness. Hameroff and Penrose have subsequently collaborated on developing and refining the theory of Orch OR, by which quantum computations in microtubules influence neural firing and by extension constitute the neural manifestation of consciousness \cite{hameroff2014}. Microtubules in general have elicited interest in various researchers attempting to model quantum effects in the brain.

\subsubsection{Microtubules}

Microtubules are formed by the polymerisation of tubulin dimers, which consist of $\alpha$ and $\beta$ tubulin proteins. Tubulin dimers first form longitudinal protofilaments; 13 of these protofilaments then form a microtubule with diameter of approximately 25 nm \cite{hameroff2014}. Microtubules form part of the cytoskeleton of eukaryotic and some prokaryotic cells and contribute to cellular shape and structure. They have a variety of functions. They are integral to cell division, forming the spindle apparatus that mediates the division of chromosomes into daughter cells. Microtubules also act as tracks along which motor proteins move cellular constituents within the cell \cite{hameroff2014,meunier,shelley}. Whereas microtubules are present in all eukaryotic cells the theory of Orch OR is focused on microtubules in nerve cells and in particular those found in the dendrites and cell body (soma) of these cells. This is because microtubules in axons and non-neural cells have a radial, regular arrangement that is arguably less supportive of information processing.  Microtubules in the dendrites and soma are less regularly arrayed, forming what Hameroff and Penrose refer to as recursive networks well suited to learning  \cite{hameroff2014,dustin}. Microtubules in non-neural cells are also dynamically unstable, able to disassemble in various ways. Microtubules in dendrites and nerve cell bodies are prevented from disassembly by microtubule associated proteins, rendering them more stable and able to encode the long-term information necessary to the theory of Orch OR \cite{hameroff2014,guillaud,craddock12}. The specific composition of tubulin has also lent strength to quantum models of neural processing due to the fact that it is partially composed of chromophores such as tryptophan, arranged in a manner similar to photosynthetic systems in plants and bacteria \cite{craddock14,toole}, which have been surmised to support coherent quantum effects. 

\subsubsection{The quantum model of Orch OR}
Hameroff and Penrose hypothesise that quantum computations encode information in microtubules and objective reduction is how this quantum information results in a classical output. The details by which Penrose's conception of quantum gravity gives rise to the objective reduction of the quantum wavefunction are beyond the scope of this review. In this case Hameroff and Penrose's 2014 review of the theory is instructive \cite{hameroff2014}. The choice of microtubules, and more specifically tubulin dimers, as the biological context in which Orch OR takes place has been motivated by various reasons. It has been suggested that the manifestation of consciousness is not axonal firing but rather the signal integration that occurs in the dendrites and cell bodies of nerve cells. This is given some support by the fact that gamma wave synchrony, which has been suggested to be the neural correlate of consciousness, is generated by dendritic-somatic integration potentials \cite{hameroff2014,hameroff2012}. As outlined above, microtubule arrangement in the dendrites and cell bodies of nerve cells is suitable for information processing, making them a good contender for the biological site of consciousness \cite{hameroff2014,dustin,craddock12}. The localisation of Orch OR in microtubules also offers a way to model a biological qubit, which is integral to the quantum nature of the theory. A qubit, the basic unit of quantum information, is a two-state system that can exist in a superposition of both states at the same time. Initially Hameroff and Penrose proposed that tubulin dimers might exist in a superposition of mechanical conformations coupled to London force dipoles \cite{hameroff2014,hameroff1996b,hameroff2007}. More recent iterations of the theory locate the quantum description firmly in the constituent aromatic rings (phenylalanine, tyrosine and tryptophan) that make up the tubulin proteins. These have pi orbital electron clouds that demonstrate spatial delocalisation, giving rise to London force electric dipoles, that can exist in superposition. While Orch OR was originally formulated with electric dipoles in mind, the authors now propose magnetic dipoles related to electron spin \cite{hameroff2014}.\\
\\
\subsubsection{Discussions around Orch OR}
This review gives only a basic overview of the theory of Orch OR as a means to introduce the idea of quantum models of the brain and their biological sites of action. There have been a number of objections raised against Orch OR, the details of which are given in detail in Hameroff and Penrose's 2014 review \cite{hameroff2014}. Those responses that deal with the structural viability of quantum effects in the brain are briefly addressed in this section. The problem of decoherence is an issue commonly cited against the application of quantum theory to biological systems. Quantum mechanics is conventionally applied to isolated systems at low temperatures whereas biological systems are typically described as being warm, wet and messy. In 2000 Max Tegmark addressed the question of decoherence in the context of Orch OR, concluding that calculated decoherence time scales of 10\textsuperscript{-13}--10\textsuperscript{-20} seconds are much too short for quantum effects to play any role in cognitive processes \cite{tegmark}. Hagan \emph{et al}. responded to this by asserting that Tegmark had based his calculations on a model that did not closely resemble the one they had proposed. After recalculation for a more accurate model they conclude that decoherence times are closer to 10\textsuperscript{-5}--10\textsuperscript{-4} seconds \cite{hagan}. With the development of quantum biology the question of whether quantum mechanics can contribute any meaningful insight into the functioning of biological organisms has shifted towards a positive answer. In particular the study of photosynthesis has revealed that the efficacy of energy transfer and charge separation might be enhanced by quantum effects \cite{engel,brixner,vangrondelle,schlau,panit,collini}. Thus the decoherence argument against Orch OR perhaps holds less weight than other evidence contradicting the theory.\\
\\
There has been some argument over the lattice arrangement of tubulin dimers in microtubules, which can be of two types: A and B. The predictions of Orch OR favour type A whereas actual mouse brain tissue points to the predominance of type B. Proponents of Orch OR argue that this does not necessarily discount the theory as quantum effects may only occur in that fraction of microtubules that are geometrically suitable \cite{hameroff2014,mandelkow}. Objections by McKemmish \emph{et al}. include the proposed conformational switching of the tubulin proteins that constitute the biological qubit. They note that the theory demands significant changes in tubulin structure and that any processes that might drive these conformational changes would be metabolically expensive \cite{mckemmish}. More recent reformulations of Orch OR do not require such extreme conformational switching and the different states necessary for the qubit can be achieved through superposed electric or magnetic dipoles in aromatic rings \cite{hameroff2014}. McKemmish \emph{et al}. also note that the electrons in a single aromatic ring cannot exist in superposition states, being completely delocalised. This can be incorporated by considering the electron clouds of two or more rings \cite{hameroff2014}. Reimers \emph{et al}. take issue with the possibility of strong Fr\"ohlich condensation in microtubules \cite{reimers}. They explain megahertz coherence demonstrated in microtubules by Pokorn\'y \cite{pokorny} as being due to weak classical Fr\"ohlich condensation. They go on to suggest that as Orch OR involves coherent strong Fr\"ohlich condensation the theory is flawed \cite{reimers}. Hameroff and Penrose respond by citing experimental evidence of the discovery of gigahertz, megahertz and kilohertz resonance in single microtubules. They argue that the experimental evidence argues strongly in the favour of Orch OR \cite{hameroff2014,sahu,sahu2}. Other criticisms have been aimed at the objective reduction part of the theory, and the way in which this translates into consciousness. These have been addressed in detail by Hameroff and Penrose \cite{hameroff2014}.\\
\\
\subsubsection{Microtubules as a site for quantum effects}
While Orch OR theory is not without controversy, the suitability of microtubules as a site for quantum effects has given rise to a number of related approaches. It seems probable that neuronal microtubules are indeed implicated in consciousness and cognition. This is supported by the fact that certain chemicals that influence both consciousness as well as cognitive function, such as general anaesthetics and antidepressants, involve microtubules \cite{hameroff2014,emerson,bianchi}. The proposed quantum action and alternative biological sites of processes related to both anaesthetics and antidepressants will be discussed in the sections B and C. Craddock \emph{et al}. suggest an additional way in which microtubules might play a quantum role in neural processing. They take, as a comparative example, evidence of quantum beats in the light-harvesting complexes of plants and bacteria. They then suggest that the tryptophan residues present in the tubulin proteins that constitute microtubules are structurally and functionally capable of supporting the possibility of quantum coherent energy transfer \cite{craddock14}. The details of this research will be discussed further in the context of electron transfer processes in section D.
\subsection{\label{sec:citeref} Quantum models of general anaesthesia}
One of the ways in which we might understand the mechanism of consciousness or, less ambitiously, the details of neural processes, is by looking at chemicals that disrupt these processes. As Luca Turin puts it, `the only thing we are sure about consciousness, is that it is soluble in chloroform' \cite{rinaldi}. To this end the study of general anaesthetics has offered some way of structuring research into quantum neural processing. There are a few theories as to how quantum effects are implicated in the action of general anaesthetics. One of these follows from the investigation of possible quantum effects in microtubules, initiated by Hameroff and Penrose's Orch OR theory \cite{craddockGA}. Another looks at spin changes in anaesthetised fruit flies \cite{turinGA}. Both of these are preoccupied with the possibility that anaesthetic action disrupts electronic activity, a theory that first made its appearance in the 1980s \cite{hameroffGA1,hameroffGA2,hameroffGA3,hameroffGA4}. There is also some suggestion that the nuclear spin of anaesthetic molecules might influence their efficacy \cite{liGA}.

\subsubsection{The action of anaesthetics in tubulin proteins}
Craddock \emph{et al}. argue that understanding general anaesthetics as having a network \cite{voss} or synaptic based effect \cite{richards} does not explain how anaesthetics inhibit the cognitive abilities of simple single celled organisms such as slime moulds \cite{perouansky,keller}. The fact that this cognition is linked to cytoskeletal microtubules points to a possible site for the action of general anaesthetics \cite{craddockGA,craddockbind}. There is some uncertainty concerning the exact biological location targeted by anaesthetics. The Meyer-Overton rule links the potency of anaesthetics to their increased lipid solubility. Further research suggests that anaesthetics act in lipid-like hydrophobic regions in proteins \cite{craddockGA}. Anaesthetics bind to a number of membrane and cytoplasmic proteins \cite{eckenhoff1,eckenhoff2,sonner}, Craddock \emph{et al}. propose that tubulin, the protein subunit of microtubules, seems the most likely due to the fact that gene expression after exposure to anaesthetic compounds is concentrated on microtubule dependent functions \cite{craddockGA,panGA}. The quantum mechanism by which anaesthetics operate in microtubules is based on a number of theoretical observations. Building on research that tryptophan rings arranged favourably in tubulin proteins can act as quantum channels supporting coherent electron dynamics in a manner similar to photosynthesis \cite{craddock14,craddockbeats,celardo}, Craddock \emph{et al}. argue that anaesthetic gas molecules binding in these channels inhibit quantum effects and disrupt coherent energy transfer and that this is responsible for the effects of general anaesthetics on consciousness \cite{craddockGA,craddockbind}. In a later paper Craddock \emph{et al}. expand their theory to the prediction of anaesthetic potency. They argue that anaesthetic and related gases change collective terahertz dipole oscillations in a way that predicts the potency of their action \cite{craddockTHZ}.

\subsubsection{Quantum spin and general anaesthetics}
In his early research relating to quantum biology, Turin hypothesised that the sense of smell, understood classically as a lock-and-key mechanism, might instead be better explained by quantum theory. More recently he has addressed the common thread between the very different molecules that act as general anaesthetics. Turin's theory is that anaesthetic molecules perturb electron currents in their target proteins \cite{turinGA}. The molecules that comprise the group of general anaesthetics range from the structurally simple noble gas xenon to the much more complex molecule alfaxalone, a range that includes numerous other chemicals without any apparent similarities. It is this structure-function anomaly that is motivation to look at the underlying physics of anaesthetics \cite{albert,turinGA}. Being that it is structurally the simplest of the anaesthetics, more than one attempt has been made to understand the anaesthetic action of xenon.\\
\\
In a recent paper Li \emph{et al}. examine the differing anaesthetic effects of xenon isotopes. Xenon has nine stable isotopes. Seven of these have zero nuclear spin but xenon 129 has a nuclear spin of $\frac{1}{2}$ and xenon 131 of $\frac{3}{2}$ \cite{liGA}. In their experiment Li \emph{et al}. compare the loss of righting reflex in mice, which correlates with consciousness, under the influence of the different isotopes. In order to exclude electronic effects they also calculate the polarisabilities of the different isotopes, finding them to be undifferentiated. Their results demonstrate that xenon isotopes with non-zero nuclear spin have a lower anaesthetic effect. As the results do not show a correlation between anaesthetic effect and atomic mass they conclude that the differences must depend on the value of the nuclear spin. Although the details of the mechanism are not clear the authors hypothesise that as spin half particles have been shown to be better suited to quantum entanglement, perhaps entanglement promotes consciousness in opposition to the effects of anaesthetics \cite{liGA}. Their results recall research done by Fisher \emph{et al}. into a possible mechanism by which neural entanglement might proceed \cite{fisher1,fisher2,fisher3}. Of particular interest here perhaps are the isotope-dependent effects of lithium, another drug that alters conscious experience.\\
\\
In order to investigate whether quantum physics plays any common role in their mechanism of action Turin \emph{et al}. also focus on the physics of the xenon atom. In 1989 IBM used a scanning tunnelling microscope to manipulate 35 xenon atoms on a nickel surface into spelling out the IBM logo; where the xenon atoms extend the conducting surface of the nickel \cite{eigler}. In a similar manner, Turin \emph{et al}. suggest that xenon atoms extend the highest occupied molecular orbit (HOMO) of the protein they interact with, facilitating electron transfer, and that this effect is common to the other molecules that act as general anaesthetics \cite{turinGA}. They test this hypothesis with an electron spin resonance experiment using anaesthetised fruit flies, looking for increases in spin caused by general anaesthetic. Discounting free electron spin signals from melanin pigments the results of the experiment lead them to conclude that general anaesthetics cause a change in spin. Further verification for the hypothesis is offered by the fact that spin changes differ for anaesthetic resistant flies. In addition to this, using density functional theory they demonstrate that chemicals which display similar but less exaggerated central nervous system effects to anaesthetics, also extend the highest occupied orbital momentum but to a lesser degree \cite{turinGA}.\\
\\
While electron spin changes under the influence of general anaesthetics points towards a possible quantum mechanism, there is still some scepticism as to whether the experimental evidence is sufficient proof of principle. Turin \emph{et al}. themselves note that the spin changes could be due to melanin. Although they checked their results against flies deficient in one of three possible melanins, they concede that particularly neuromelanin, the grey in grey matter, might play a role \cite{turinGA}. This is potentially interesting in the context of recent research investigating the quantum role of neuromelanin facilitated electron transport in sections of the brain \cite{rourk}. Turin \emph{et al}. also document the fact that the experiment was performed under distinctly unphysiologic conditions. In order for the flies to remain immobile and not disturb the readings they were kept at 6 C. Although some temperature variation was tested this was only between 2 and 10 °C. In addition to this the flies were kept in anoxic conditions to isolate the effects of anaesthetic gases from those of oxygen. The potential role of oxygen in the experimental results points to the involvement of respiration, more specifically the movement of electrons in the electron transport chain of mitochondria \cite{turin18}. Turin \emph{et al}. cite a number of instances in which it has been shown that mitochondria are involved in the action of anaesthetics \cite{turinGA,kater,cohen,modelli}. In a more recent collaboration with Turin, Gaitanidis \emph{et al}. invite speculation on an unpublished preliminary report of spontaneous radiofrequency emission from fruits flies subjected to a magnetic field. They suggest that these emissions originate from the nervous system due to the fact that they stop under the influence of chloroform anaesthetic and that they are related to spin-polarised electron currents in cells. The authors note that these cellular currents could reflect mitochondrial metabolism but that the variable signal and reaction to anaesthetic suggests some more complex biological activity connected to neuronal activity \cite{gaitanidis}.

\subsubsection{The outlook on quantum models of general anaesthesia}
While the comparison of electron dynamics in photosynthetic and microtubule proteins is largely theoretical the authors suggest its experimental verification is possible via two-dimensional electronic spectroscopy, a technique well established in the analysis of photosynthetic systems \cite{futureQB}. They propose that quantum beating in tubulin will be altered under the influence of anaesthetics and that this is a measurable phenomenon \cite{craddockGA}. While this has yet to be attempted there has been some recent experimental evidence that supports the possibility that quantum effects play a role in the action of anaesthetics. Using various types of spectroscopy, Burdick \emph{et al}. report that, in contrast to nonhalogenated ethers, halogenated ethers interact with entangled photons of specific wavelength \cite{burdick}.
\subsection{\label{sec:citeref} Neurochemical binding and activation mechanisms}
For the purposes of this review the term neurotransmitter is used with respect to a variety of neurochemicals that fall into this category, binding to neuroreceptors and modulating neural signalling. Neurotransmitters emitted by one nerve cell bind to receptors on an adjacent nerve cell and facilitate the opening of ion channels. This is fundamental to the generation of action potentials and disruption of this process is believed to contribute to mental illnesses \cite{pan,nutt}. Although this review focuses on the action of neurotransmitters it has been suggested by Vaziri \emph{et al}. that the mechanism by which ion channels allow the selective transmission of ions might also not be strictly classical, showing evidence of quantum coherence \cite{vaziri}. The binding action of neurotransmitters is conventionally understood as a lock and key mechanism whereby the shape of the neurotransmitter matches its specific receptor \cite{fischer,tripathi}. The lock and key or docking mechanism is implicated in a number of biological processes: neurotransmission; the action of enzymes; olfaction; DNA binding, all depend to some extent on lock and key theory \cite{fischer,mitsuda,reese,samee}. Despite the success of the theory however, an alternative view suggests that something more than molecular shape might be necessary to explain olfaction and, more recently, neurotransmission.
\subsubsection{\label{sec:citeref}Lock and key or quantum vibration}
Neurotransmitters are a class of molecules that bind to G protein-coupled receptors (GPCRs). GPCRs are a group of receptors that, upon detection of appropriate molecules known as ligands, activate signalling pathways in eukaryotic cells. GPCRs play an important role in medical innovation, as targets for drug action \cite{rask}. In addition to neurotransmitters, there are a number of biological molecules that bind to GPCRs; these include odourants \cite{gehrckens}. Olfaction is classically understood as depending on the respective shapes of the chemical/receptor pair. However, an alternative vibrational theory of olfaction was first developed by Dyson as long ago as the 1930s \cite{turinVTO,dyson}. In his 1996 paper Turin suggested that the vibrational frequency of a given odourant contributes to quantum tunnelling at the receptor \cite{turin96}. In 2011 Franco \emph{et al}. presented experimental evidence in support of the theory. In particular they claimed that fruit flies could differentiate between deuterated odourants \cite{turinFF}. There is also some evidence that more complex species such as lake whitefish and the American cockroach can differentiate between isotopes of amino acids and pheromones \cite{hoehn,hara,havens}. Horsfield \emph{et al}. document a number of experiments that test the theory on both the behavioural and physiological level \cite{turinVTO}. Deuterated odourants are a useful means to test the vibrational theory of receptor activation due to the fact that replacing hydrogen by deuterium adds vibrational modes, for instance the 2150 $cm^{-1}$ vibration of the carbon-deuterium bond \cite{turinVTO,klika, paoli}. However, there has also been some recent discussion as to the fact that a differential olfactory receptor response between undeuterated and deuterated odourants could in fact be due to a minute contaminant in one of the samples. Paoli \emph{et al}. conclude that although their results do not prove the vibrational theory of olfaction they also do not disprove it, merely calling for caution in experimental approach \cite{paoli}. While there is little verification for a vibrational olfactory mechanism in mammals, a number of the experiments support the theory for the case of insects. It should be noted, in a discussion of GPCR mechanisms, that insect olfactory receptors differ from mammalian G-protein related olfactory receptors and thus experimental results may not be generalisable \cite{turinVTO}.\\
\\
While the theory remains controversial it has recently been re-examined in the context of neurotransmission. A number of different neuroreceptors have been investigated. Adenosine receptors are rhodopsin-related G-protein coupled receptors that bind the neuromodulator adenosine, but can also bind a variety of other agonist and antagonist molecules \cite{chee1}. The stimulating action of caffeine, for instance, is due to its antagonistic binding to adenosine receptors. Because GPCRs are such an important target for pharmaceutical intervention the classification of molecules that act as agonists or antagonists is a well-developed research field. Molecules can be classified using various molecular descriptors which include information from molecule structure, topology and geometry to dipole moment, electric polarisability and electrostatic potential  \cite{chee1}. Following on from the fact that adenosine receptors and olfactory receptors are both class A GPCRs, Chee and Oh present research that tests whether vibrational frequency might also be an effective molecular descriptor. They suggest that classifying ligands by vibrational frequencies is an effective way of discriminating agonist from antagonist \cite{chee1}. Chee \emph{et al}. then refine this research using a machine learning approach and conclude that selected features of molecular vibration allow for ligand classification of adenosine receptor agonism \cite{chee2}.\\
\\
Whereas Chee \emph{et al}'s research investigates adenosine receptors Hoehn \emph{et al}. focus in particular on the neurotransmitter serotonin and its receptors \cite{hoehn,hoehn2}. Although it has a number of other functions in biological systems serotonin is perhaps most well known for the role it plays in mood. A widely prescribed class of antidepressants, the SSRIs (selective serotonin reuptake inhibitors) target this neurotransmitter \cite{lin}. In their research Hoehn \emph{et al}. address the possibility that serotonin neurotransmission might utilise vibration assisted inelastic tunnelling effects. Using inelastic electron tunneling spectroscopy (IETS) theory that was first developed to understand olfaction they investigate the tunnelling spectra of endogenous and non-endogenous agonists that bind to the serotonin receptor. Non-endogenous agonists of this receptor include LSD (lysergic acid dimethylamide), DOI (2,5-dimethoxy-4-Iodo-amphetamine) and other psychedelic phenethylamines \cite{hoehn}. Their results suggest that the serotonin molecule shares an inelastic electron tunnelling spectral peak with other agonists that activate the serotonin receptor. Although the lock and key mechanism has been very successful in modelling certain aspects of the action of signalling proteins one of the ways in which it falls short is the prediction of agonist potency. In addition to the shared spectral peak of related agonists Hoehn \emph{et al}. also report that the intensity of this peak might be used as a predictor of agonist potency \cite{hoehn}.
 \subsubsection{\label{sec:citeref}Experimental verification of the vibration assisted tunnelling hypothesis}
Experimental verification of the vibrational theory of neuroreception follows the lead of Franco \emph{et al}. in the context of olfaction, where the effects of deuterated odourants are investigated \cite{turinFF}. Hoehn \emph{et al}. report the results of an experiment to test the feasibility of their theoretical approach by measure of receptor affinity and activation. Once again, agonists of the serotonin receptor were chosen, specifically 2,5-dimethoxy-4-iodoamphetamine (DOI) and N,N-dimethyllysergamide (DAM-57). However the experiment failed to confirm the vibrational theory of neuroreception as they report that selective deuteration had no influence on binding affinity or activation \cite{hoehn2}.\\
\\
In another study regarding the binding mechanism of the neurotransmitter histamine, authors Kr\v{z}an \emph{et al}. report a significant distinction in the binding patterns of histamine and its deuterated counterpart, as well as for other agonists of the histamine receptor \cite{krzan}. In particular they found that deuterating histamine increased its binding affinity. They discuss this in the context of the vibrational theory of olfaction and GPCRs more broadly. Instead of confirming the theory they propose an alternative reason for the fact that experiments suggest that animals can differentiate between deuterated odourants, concluding that it is a nuclear quantum effect governed by differences in the strength of hydrogen bonds before and after deuteration \cite{krzan}. In contrast to this, theoretical work comparing the structure-activity/vibration-activity relationship of the histamine receptor and its various ligands suggests that molecular vibration does play some role in ligand function. In the study 47 ligands that bind to histamine receptors were investigated using a computational approach to molecular vibration. This led to the conclusion that the many varying agonists and antagonists can be to some extent classified by their molecular vibrations \cite{oh}.
\subsubsection{\label{sec:citeref}The functional mechanisms of GPCRs in general}
The lack of experimental verification for vibrational quantum effects in the context of central nervous system GPCRs prompts Hoehn \emph{et al}. to suggest that either olfactory receptors function differently from other GPCRs or the vibrational theory of receptor activation is wrong. Gehrckens \emph{et al}. address the latter criticism in a recent preprint, in which they use density functional theory to investigate the electronic structure of rhodopsin \cite{gehrckens}. They choose to look at rhodopsin rather than olfactory GPCRs due to the fact that a high resolution structure of rhodopsin is available for research purposes. Rhodopsin also shares important features with olfactory receptors and is considered the evolutionary ancestor of GPCRs \cite{palczewski,krishnan}. The main aim of their research is to demonstrate a mechanism for electron transfer in olfactory receptors. They go on to identify a tryptophan donor and zinc acceptor that would indeed allow electron transfer in rhodopsin, although they also emphasise the fact that this capacity for electron transfer does not necessarily play a role in the functionality of rhodopsin. In the context of whether this effect might be generalisable to GPCRs they conclude that an electron transfer mechanism might have been exploited by the offshoots of rhodopsin, in particular olfactory receptors. They do however suggest that in this sense olfactory receptors may be unique in that they are relatively non-specific, a requirement made necessary by the novelty and variety of odourants that will bind to them. Neurotransmission, they argue, depends on a binding that is much more specific. On the subject of neurotransmitters, they hypothesise that the anomalous binding structure of the antagonist metitepine to serotonin related receptors can be explained if the metitepine is in the form of a radical, having gained an electron in binding to the receptor. This mechanism is in line with the idea that GPCRs are electronic devices rather than simply lock-and-key \cite{gehrckens}.\\
\\
Both theoretical and experimental results do not offer any clear conclusions with respect to the vibrational theory of neurotransmission for adenosine, serotonin and histamine receptors and their related ligands. Further research is necessary to clarify the viability of the approach and its specific relation to the different actions of neurotransmitter binding affinity and activation capability \cite{chee2}. Understanding the mechanism or mechanisms of ligand-GPCR interaction is particularly important due to the fact that GPCRs are a major drug target associated with one third of all pharmaceuticals \cite{chee2,rask}.
\subsection{\label{sec:citeref} Alternative signalling processes and biophotons}
It is perhaps interesting that GPCRs, in addition to binding with molecules such as neurotransmitters and odourants, also bind to photons \cite{wacker}. There has recently been some suggestion that cellular communication and even possibly neural signalling might make use of biophotons in addition to more well established mechanisms such as neurotransmitters \cite{fels,rahnama,kumar}. Biophotons, are spontaneous ultra-weak photons in the near-UV to near-IR spectral range produced by biological systems, in particular through oxidative processes in mitochondria \cite{kumar,cifra}. As noted in a recent paper on the subject, it is known that biophotons are produced in brains \cite{kumar}. Correlations have been found between biophoton intensity and neural activity as well as oxidative dysfunction of neural cells in rat and mouse brains \cite{kobayashi,isojima,kataoka}. There has even been some attempt to explain human intelligence in the context of biophoton emission. Following on from a study that demonstrated that glutamate, a key neurotransmitter, can mediate biophoton production \cite{tang}, Wang \emph{et al}. examine the spectral characteristics of biophoton emission in a range of species. They conclude that there is a correlation between higher order intelligence and the spectral redshift of biophotons emitted by sample brain slices. This redshift increases in the order of frog, mouse, chicken, pig, monkey, and human, which reflects with some accuracy the phylogenetic tree, although they concede that there is no completely objective means to measure intelligence across species \cite{wang1}. Their argument has met with some scepticism, with Salari \emph{et al}. responding that the experimental results insufficiently support the hypothesis and that without a mechanistic connection between spectral shift and intelligence, their correlation is more likely to be coincidence \cite{salari}.\\
\\
Biophotons have been implicated in cellular signalling in plants, bacteria and even kidney cells. Following from this Sun \emph{et al}. present experimental evidence that nerve cells can also conduct biophotons, and that this effect can be inhibited by the application of a neural conduction block anaesthetic \cite{rahnama,sun}. In their attempt to model the mechanistic details as to how biophotons are utilised in information transfer, Kumar \emph{et al}. argue that neurons are well suited to photonics \cite{kumar,zarkeshian}. They cite evidence that nerve cells contain possible photon sources from mitochondrial respiration or lipid peroxidation, as well as photon detectors such as centrosomes and chromophores \cite{kumar,zhuravlev,tuszynski,albrecht,kato}. They then hypothesise that myelin-coated axons of nerve cells act as waveguides for biophotons and that this might facilitate quantum effects such as entanglement \cite{kumar}. In order to demonstrate proof of principle the authors develop a theoretical model of light guidance in axons, solving the three dimensional electromagnetic field equations in the relevant context. They investigate a number of limiting factors and potential pitfalls and conclude that axons are mechanistically viable as waveguides. They go on to suggest various experiments in which the different aspects of their hypothesis might be put to the test \cite{kumar,zarkeshian}.
\subsubsection{\label{sec:citeref}Quantum effects in electron transfer}
In a review that addresses the possibility of quantum neurobiology, Jedlicka suggests that cognition, typified by complex signal processing and integration, can potentially be thought of as occurring outside of strictly neural systems and that the quantum information processing abilities demonstrated by plants and bacteria could be useful to the study of cognition in higher animals \cite{jedlicka}. Thus far, the theoretical and experimental advances made in understanding the role of quantum effects in biological systems have been primarily to do with photosynthesis, in particular light harvesting and energy/charge transfer processes \cite{futureQB}. Many of the theories of quantum effects in the context of neural processes focus on electron dynamics. As noted by Toole \emph{et al}. coherent energy transfer in photosynthesis is less a unique feature of photosynthetic systems than it is the result of the specific arrangement of chromophores within a protein \cite{toole}.\\
\\
Kurian \emph{et al}. propose that biophoton production in mitochondria is absorbed and channelled via resonant energy transfer by co-localised microtubules. They argue that neurodegenerative diseases related to compromised microtubule networks could result from ineffective channelling of biophotons into signalling or dissipation. Experimental evidence shows that microtubules undergo organisational changes after exposure to photons, particularly in the absorption range of tryptophan and tyrosine \cite{kurian}. As previously noted, a related paper presents a computational approach to quantum coherent energy transfer in microtubules \cite{craddock14}. The authors argue that similarly to the arrangements of chromophores in light harvesting photosynthetic complexes, the tubulin proteins that constitute microtubules have appropriate arrangements of chromophores such as tryptophan. They conclude that it is feasible that tubulin proteins could support coherent energy transfer in microtubules \cite{craddock14,craddockbeats}. In a very recent related paper Celardo \emph{et al}. extend the similarities between photosynthetic antenna complexes and microtubules by demonstrating that tryptophans in microtubules can theoretically exhibit superradiant excitonic states \cite{celardo}.\\
\\
The possible role of tryptophans in electron transfer processes has also been previously addressed in this review with respect to the action of neurotransmitters in binding to and activating GPCRs. In their investigation of rhodopsin, Gehrckens \emph{et al}. outline a mechanism by which a tryptophan donor and zinc acceptor facilitate electron transfer \cite{gehrckens}. Tryptophan seems particularly interesting with respect to neural processes as it is the precursor to serotonin, an important neurotransmitter. This connection prompts Tonello \emph{et al}. to hypothesise that the structure of consciousness emerges in a manner analogous to the serotonin dependent movement of plants towards their source of energy \cite{tonello}. In neural processing this is extrapolated as biophoton harvesting by tryptophans and concomitant serotonin-mediated neural communication and plasticity \cite{tonello} In a later paper Tonello \emph{et al}. expand on this idea with the hypothesis that the gastrointestinal-brain axis in higher animals might have evolved from the root-branch axis of plants, and that light plays an important role in both systems \cite{tonello2}.\\
\\
\begin{figure}
	\begin{center}
		\includegraphics [scale = 1.2]{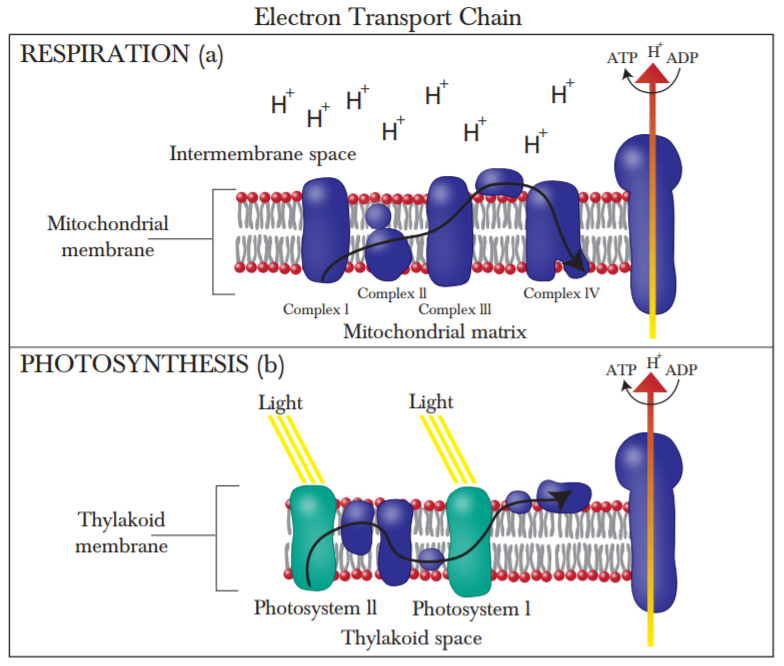} 
		\caption{Simplistic comparison of the electron transport chains of respiration and photosynthesis. Electrons are transported through a series of complexes or photosystems, which results in the creation of a proton gradient. This is used to drive the production of ATP. While the specific complexes differ between chains they share some features relevant to the possibility of quantum processing. The presence of chromophores, for instance, which arguably support coherent energy transfer. While the interaction of light with the photosynthetic electron transport chain has been studied in detail, analogous effects in mitochondria are not well understood. Oxidative processes in the electron transport chain of mitochondria are thought to be the main source of biophotons. The function or target of these photons, however, is less clear. It has been proposed that microtubules close to mitochondria absorb and channel biophotons in a quantum coherent manner \cite{kurian}. Mitochondria themselves also demonstrate the capacity for photon absorption, in particular complex IV in the electron transport chain, which is the target of therapeutic application of infrared light to treat disorders of the brain \cite{salehpour,hamblin2,karu}.}\label{fig1_schematic}
	\end{center}
\end{figure}
While these ideas remain for the moment largely theoretical, much of the theory of electron transport and resonant energy transfer has focused on the involvement of microtubules. Alternatively, the role of biophoton use and energy transfer in neural processes could be furthered by a closer look at the electron transport chain of mitochondrial respiration. As previously noted, mitochondria are already implicated in the action of general anaesthetics \cite{turinGA,kater,cohen,modelli}. They are also the primary production site of biophotons. They might serve as an alternative biological location in which quantum effects could be investigated. Research suggests that the arrangement of chromophores in bacteria allows for quantum coherence in their photosynthetic processes \cite{baghbanzadeh}. Similarities between the DNA of bacteria and eukaryotic mitochondria has led to the theory that mitochondria are descended from bacteria, though recent research suggests they are less directly related \cite{gray,wangMIT,harish,martijn}.\\
\\
While the ancestry of mitochondria remains uncertain their similarity to bacteria could arguably manifest in biological structures that support quantum effects. Both photosynthesis and mitochondrial respiration make use of an electron transport chain, which powers the charge gradient necessary for ATP synthesis. A group of researchers from the Quantum Biology and Computational Physics research group at the University of Southern Denmark are already investigating proton-coupled electron transfer in the cytochrome bc1 complex in photosynthetic bacteria and higher organism cellular respiratory systems. The group has published a number of papers motivated by the fact that malfunctions in the bc1 complex lead to many different diseases. They also hope to better understand photosynthesis in order to optimise energy conversion research \cite{husen,salo,barragan}. The reduction and oxidation of ubiquinone is essential to the electron transfer reactions in cytochrome bc1 and the concentration of ubiquinone is particularly high in the high energy consumption environment of the brain \cite{xia,salehpour}. There is some evidence that transcranial photobiomodulation, the application of red or near-infrared laser light to the cranial area, is an effective treatment for various brain disorders \cite{hamblin,hamblin2}. It has also shown promise in treating depression \cite{changlight} and when coupled with ubiquinone supplementation \cite{salehpour}. Research suggests that, among other effects, photobiomodulation can improve attention, memory, executive function and even rule-based learning \cite{barrett,blanco1,blanco2}. The mechanism of this effect is still not completely clear but the site of the effect is most likely the mitochondrial electron transfer chain, in particular the chromophores in the different complexes that constitute the chain \cite{salehpour,hamblin2,karu}. There are a number of these chromophores, including tryptophan, which plays a central role in the various theories that espouse quantum effects in neural processing. Given that the role of quantum effects in light harvesting, electron and charge transfer has contributed to a better understanding of elements of the electron transfer chain in photosynthesis, this knowledge might be fruitfully applied to the mitochondrial electron transport chain.

\subsubsection{\label{sec:citeref}Magnetic field effects and the brain}
The importance of tryptophans in energy transfer is evident in another well established field of quantum biology: the avian compass. A leading theory of bird migration suggests that birds utilise the earth's magnetic field as a guiding tool by means of the radical pair mechanism \cite{schulten78,wandw,horerodgers}. The radical pair mechanism can be summarised in three main steps. First, a photon incident on a molecule causes electron transfer and pair formation. Second, the radical pair, originally in singlet spin state, interconverts between singlet and triplet state under the influence of the nuclear hyperfine and geomagnetic Zeeman effects. And finally spin-dependent recombination leads to some signalling state that the bird interprets as a spatial directive \cite{horerodgers}. It is generally accepted that the molecule in which this occurs is the blue light activated flavoprotein cryptochrome \cite{ritz10,pinzon18,liedvogel14,mouritsen18}. Theoretical research, backed by spectroscopic evidence, suggests that light activated electron transport occurs between flavin adenine dinucleotide (FAD) and tryptophan residues \cite{horerodgers}.\\
\\
Although it is well known that a number of species have a functioning magnetic sense, humans have not yet been added to that list. In a recent paper, Wang \emph{et al}. present the results of an experiment that demonstrates the effects of earth strength magnetic fields on the human brain \cite{wang2}. The authors report that magnetic field changes result in a decrease in amplitude of alpha frequency (8-13 Hz) brain waves, an effect normally associated with the brain's processing of external stimuli. They conclude that the effect is likely to be due to ferromagnetism rather than the radical pair mechanism. This is because the effect is dependent on the polarity of the field, for subjects in the Northern Hemisphere the alpha wave response only occurs for horizontal rotations if the static component is directed upwards \cite{wang2}. While this would appear to exclude the radical pair compass which is dependent on inclination, the corresponding effects would perhaps need to be confirmed for subjects in the Southern Hemisphere. It has been suggested, in the context of avian migration, that birds employ both ferromagnetism and the radical pair mechanism for navigation \cite{wandw,wiltschkofixed,kishkinev} and this could also be true of humans' magnetic sense. The question being whether there is any evidence that might link magnetic effects in humans with the radical pair mechanism.\\
\\
Cryptochrome, the proposed site of the avian compass, is also present in humans \cite{hsu96}. Foley \emph{et al}. use a transgenic approach to show that human cryptochrome can act as magnetosensor in the magnetoreception of fruit flies \cite{foley}. It is also potentially interesting that is it alpha waves that are effected by the magnetic field changes. Van Wijk \emph{et al}. present evidence that biophoton fluctuation is correlated with the strength of alpha wave production, where they measure biophoton production in terms of the fluctuations of reactive oxygen species (ROS) \cite{rahnama,vanwijk}. The production of ROS and how this relates to the radical pair mechanism has been the motivation for various papers published in the field of quantum biology. Marais \emph{et al}. outline a quantum protective mechanism in photosynthesis. They hypothesise that the high spin iron in the reaction centres of photosystem II of the electron transport chain exerts a magnetic field effect that reduces the triplet yield of radical pairs that are formed during electron transfer. Triplet states are instrumental in forming ROS which are toxic to living cells \cite{schweitzer,marais}. In another paper, Usselman \emph{et al}. show how yields of ROS in live cells are altered by radical pair dynamics, in particular coherent singlet-triplet mixing under the influence of oscillating magnetic fields at Zeeman resonance \cite{usselman}. If ROS are correlated with alpha wave production \cite{rahnama,vanwijk} and ROS yields are altered by the Zeeman effect \cite{usselman,usselman2}, then this might point to a mechanism that explains the influence of geomagnetic fields on alpha waves in the human brain.\\
\\
Reactive oxygen species participate in cellular signaling \cite{wangROS} and have been implicated in aging and numerous diseases including mental conditions such as depression and schizophrenia \cite{shigenaga,stadtman,michel,salim,salim2,kovacic}. If humans do have physiological systems that depend on the dynamics of radical pairs then it is expedient to understand exactly how these function. The avian compass is disrupted by broadband radiofrequency electromagnetic radiation that is well below the WHO-recommended level \cite{ritzres,naturenoise}. This radiation might equally be upsetting the balance of ROS in other species and in turn leading to biological malfunction. There has been some suggestion that the radical pair mechanism is implicated in the development of cancer through circadian rhythm disruption and related ROS levels \cite{horecancer}. Though the threat of radical pair mediated carcinogenesis is debatable, being equivalent, as one study finds, to the risk of travelling some kilometres towards or away from the  earth's magnetic poles \cite{hore60}. Nevertheless, the disruption of circadian rhythms has also been linked to mood disorders and cognitive function \cite{lyell,boyce,ferguson}. Although the evidence remains contentious, there is also some documentation of the psychological effects of geomagnetic storms, which alter the earth's magnetic field for a limited period of time \cite{close,kay,ora}. The radical pair mechanism could offer a testable hypothesis as to how these effects occur \cite{close}.
\begin{figure}
	\begin{center}
		\includegraphics [scale = 0.9]{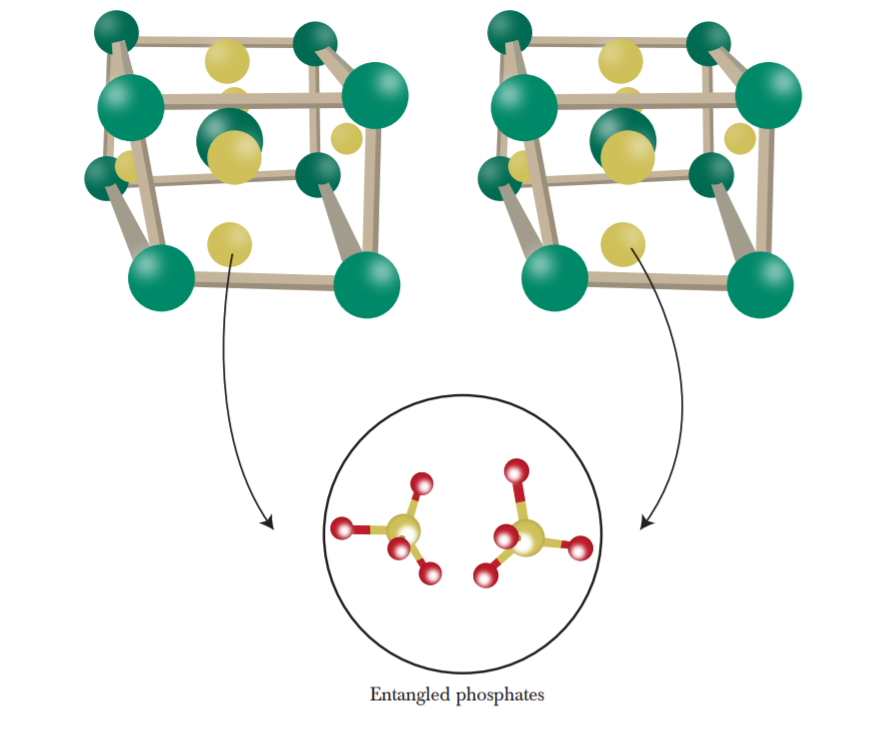} 
		\caption{A scheme for neural entanglement. The Posner molecule structure is interpreted loosely here as a distorted cube with calcium ions (green) at each vertice and at the centre, and a phosphate ion (yellow) on each of the six faces \cite{hore}. This diagram is not meant to accurately convey the exact geometry of the Posner molecule but rather give some sense of Fisher's proposed qubit. The two Posner molecules share an entangled pair of phosphate ions. When two entangled Posner molecules are spatially separated and subsequently taken up into different presynaptic neurons in the process of vesicle endocytosis, their entanglement results in the correlated release of neurotransmitters and the coordination of firing across neurons \cite{fisher1}.}\label{fig1_schematic}
	\end{center}
\end{figure}
\subsection{\label{sec:citeref}Neural entanglement}
One of the first and strongest objections to quantum models of consciousness was the phenomenon of decoherence: that the non-ideal environment of biological systems would destroy any quantum effects before they could prove useful. As initially calculated by Max Tegmark in response to Orch OR theory, the timescales on which decoherence occurs in the environment of the brain are considerably shorter than neural firing rates \cite{tegmark}. This conclusion has subsequently been challenged not simply by researchers interested in Orch OR, but also by the development of the field of quantum biology. If quantum effects play a role in photosynthesis and possibly other biological contexts, then it is not such a stretch to consider that they may play a role in neural processes. Decoherence, it has been argued, might even enhance energy transfer \cite{huelga}. While research suggests that long-lived coherence in photosynthetic systems lasts for picoseconds \cite{gsengel} and the lifetime of the radical pair mechanism is discussed in terms of milliseconds \cite{horerodgers}, it seems unlikely that coherence in biological systems extends beyond these timescales. That is until a new hypothesis concerning neural entanglement proposed coherence that could last for hours or even days.
\subsubsection{\label{sec:citeref}Entangled Posner molecules}
The hypothesis is this: that phosphorus nuclear spin might function as a neural qubit to allow for quantum processing to play a role in cognition \cite{fisher1,fisher2,fisher3}. Following on from the proposal by Hu and Wu, that consciousness is linked to quantum spin \cite{huwu}, Fisher argues that the spin-half phosphorus nucleus seems the only likely candidate for this role due to the fact that for any nucleus with spin greater than half, the electric quadropole moment means quicker decoherence due to electric field interactions in addition to the slower decoherence caused by the magnetic fields of nearby nuclei. Spin-half nuclei are thus more favourable in terms of decoherence times. Out of the various elements and ions that play an essential role in biological systems, only hydrogen and phosphorus have spin-half nuclei. Fisher goes on to describe the details of how the quantum information carried by phosphorus nuclei is created and preserved \cite{fisher1,fisher2,fisher3}.\\
\\
Phosphorus is bound into phosphate or polyphosphate ions such as pyrophosphate. Phosphate ions constitute part of the adenosine triphosphate (ATP) molecule, an organic chemical that acts as a source of energy for the many essential actions that sustain living organisms \cite{fisher1,cole}. The electron transport chain in both photosynthesis and respiration creates a proton gradient in order to drive ATP production. Fisher postulates that when adenosine triphosphate is hydrolysed to adenosine monophosphate and pyrophosphate the two phosphorus nuclei in the pyrophosphate ion will have a specific spin alignment, either a singlet or one of three triplet states. A quantum mechanical treatment of the enzyme catalysed reaction that creates two phosphate ions out of pyrophosphate demonstrates that this reaction depends on the nuclear spin state. More specifically, the enzyme conditional outcome of the reaction, where the spin dynamics of the triplet states are unfavourable, mean that the phosphorus nuclear spins in the two distinct phosphate ions will emerge in a singlet entangled state \cite{fisher1}.
\subsubsection{\label{sec:citeref}Coherence timescales for Posner molecules}
Phosphorus in the phosphate ion is surrounded by an oxygen cage, where the oxygen nuclei all have zero spin. Despite this the coherence time is still very short for solvated phosphates, approximately one second, due to the fact that hydrogen quickly binds to phosphate and the non-zero proton spin contributes to decoherence. A coherence lifetime of a second is long enough for the quantum effects to be sustained over the process of cellular diffusion \cite{fisher1}. However in order for effects such as memory storage and retrieval to utilise quantum effects, coherence lifetimes need to be significantly longer. Fisher argues that if binding with the hydrogen can be pre-empted by a spin zero cation such as calcium, then longer coherence times might be possible. He identifies a possible molecule as the Posner molecule \cite{posner}, thought to be involved in the formation of hydroxyapatite, an important constituent of bone tissue \cite{hore,wangposner,mancardi}. A Posner molecule is, simplistically, a distorted cube with a calcium ion at each vertice, a ninth at the centre and a phosphate ion on each of the six faces of the cube \cite{hore}. The entangled phosphate spins then result in entangled Posner molecules \cite{fisher1}.\\
\\
Fisher estimates that for phosphorus spins in Posner molecules coherence times could be as long as hours \cite{fisher1} or, in a later paper, 21 days \cite{fisher2}. For solvated Posner molecules the magnetic fields from protons in surrounding water molecules cause decoherence but, as Fisher argues, the rapid tumbling of Posner molecules means that the average magnetic field felt by a phosphorus nucleus will be zero. Decoherence will then only happen from residual magnetic field fluctuations and coherence times will thus be much longer \cite{fisher1,fisher2}. Player \emph{et al}. respond to this with their own calculation of coherence times for Posner molecules. They acknowledge that long lived spin states in nuclei are well accepted, with relaxation times of up to an hour being recorded \cite{hore,stevanato}. However, they argue that these extended times are partly due to experimental control of the coherent spin dynamics, a condition that Fisher does not take into account \cite{hore}. In their analysis of Posner spin dynamics they take a number of other factors into account such as dipolar and scalar couplings within Posner molecules as well as the Zeeman interaction of the phosphorus spins with the geomagnetic field \cite{hore}. With these as the dominant relaxation pathway they arrive at a relaxation time of 37 minutes, as opposed to Fisher's shortest estimation of approximately a day. They also discuss other ways in which this relaxation could take place even more quickly \cite{hore}.
\subsubsection{\label{sec:citeref}Posner molecules in their biological context}
Fisher also outlines the way in which entangled Posner molecules give rise to quantum effects in neural processes. The entanglement of the spin and rotational states of Posner molecules leads to correlations in the chemical reactions of spatially separated Posner molecules. These entangled molecules are taken into presynaptic glutamatergic neurons in the process of vesicle endocytosis. The acidic environment causes Posner molecules to bind and release calcium which stimulates exocytosis and further release of glutamate, which enhances neural firing. Thus, because the chemical reactions of Posner molecules are entangled the subsequent release of glutamate and resultant neural firing might also be considered entangled \cite{fisher1,fisher3}. Fisher outlines a number of possible experiments that might verify the various stages of the theory, most pressingly whether Posner molecules are present in bodily fluids. Whereas free-floating calcium phosphate clusters resembling Posner molecules have been found to be stable in simulated body fluids \cite{ito}, there is still some uncertainty as to whether they are present in vivo \cite{fisher1,fisher2}. Less fundamental but more interesting in the context of this review is the suggested investigation of the effects of a chemically viable replacement of the central calcium in a Posner molecule with lithium ions \cite{fisher1,fisher2}. There is some evidence that lithium, used to alleviate the symptoms of bipolar disorder, has isotope dependent effects on the behaviour of rats \cite{fisher1,sechzer}. Fisher postulates that the efficacy of lithium as a treatment for mental disorders could be due to the increased decoherence induced by the lithium nuclear spins included in the Posner molecule \cite{fisher1,fisher3}.\\
\\
\subsubsection{\label{sec:citeref}Neural qubits and quantum computing}
While the results of experiments to verify this model of entangled neural processes are yet to be completed the theory has caught the interest of researchers working in the field of quantum information theory. Halpern and Crosson apply Fisher's Posner molecule theory in the context of quantum communication, quantum computation and quantum error correction. They address how a quantum information based model of Posner molecules might contribute to the understanding of Posner chemistry and vice versa, what insights Posner molecules have for quantum information processing. They also demonstrate how entanglement can change molecular binding rates, which goes some way to supporting Fisher's neural entanglement hypothesis \cite{halpern}. The authors conclude that the quantum information based formulation of Posner molecules might be a way of framing biological Bell tests \cite{halpern}, a claim that is potentially interesting in light of recent observations of entanglement between living bacteria and quantised light \cite{marletto}.\\
\\
This is not the first time that phosphorus spin and quantum computing have crossed paths. In 1998 Kane proposed a scalable quantum computer where information is encoded onto spin half phosphorus nuclei in a substrate of spin zero silicon, ensuring the long decoherence time necessary for quantum computing. Computations are then performed through the interaction of nuclear spin with donor electrons \cite{kane}. Kane's idea has recently been reworked by Tosi \emph{et al}. into what they call the flip-flop qubit, where the combined electron-nuclear spin states of a phosphorus donor are controlled by microwave electric fields. This research, along with work done by He \emph{et al}. into engineering a viable exchange interaction between two phosphorus bound electrons has done much to advance the course of spin based quantum computing \cite{tosi,he}. The fields of quantum computing and quantum neurobiology might inform each other in other ways too. Quantum dots have been proposed as an alternative means to implement a quantum computer \cite{kloeffel}. They have also been used to model the mechanism by which anti-acetylcholine receptor antibodies contribute to the neuromuscular disorder myasthenia gravis \cite{chilee}. Even more recently, graphene quantum dots have been shown to prevent and even undo the protein clumping of neurons in the brain that leads to Parkinson's disease \cite{kim}. A separate study study demonstrated similar results for Alzheimer's disease \cite{liu}.
\section{\label{sec:level1}Conclusion}
The field of neurophysics already attests to the fact that the study of the brain borrows from ideas across the spectrum of physics: electricity and magnetism, mechanics, thermodynamics, optics \cite{lynn}. It is perhaps not completely surprising that quantum physics might contribute too. While this review outlines working theories as to how quantum effects might be implicated in neural processes, the research remains largely theoretical. Although some experimental evidence points to the validity of certain of these theories, conflicting results mean it is difficult to draw any strong conclusions. However, many of the authors cited in this review suggest ways in which their theories might be put to the test experimentally. As already suggested by Craddock \emph{et al}. advances in experimental techniques such as two-dimensional electronic spectroscopy, which has been successfully applied to understanding the quantum nature of photosynthesis, might yield similar success in the context of energy transfer processes that impact neural signalling \cite {craddockGA}. A number of the hypotheses included in this review rework, to some extent, ideas already established with respect to other biological systems. Coherent energy transfer in photosynthesis is reimagined in the tryptophan rings of neural microtubules. The vibrational theory of olfaction is reapplied to the binding action of neurotransmitters. And while Fisher's Posner qubits depend on nuclear spin, unlike the radical pair model developed with respect to avian migration, their entangled spin dynamics mediate the outcome of signalling processes in a potentially analogous fashion. As it stands, the future of quantum neurobiology is the future of quantum biology more generally; progress made in either will further the other. The questions are then technical. How exactly to apply experimental techniques to the specific neural context. How to refine, for example, experiments that measure the differing effects of isotopes without the confounding influence of contaminants. This mechanistic focus would seem to ignore the question of consciousness and how it emerges from these underlying neural mechanisms. Lynn \emph{et al}. describe the future of brain network research, albeit in a classical capacity, as a cross-scale approach that links the microscopic to the macroscopic: protein reactions in neurons to synaptic connectivity to brain region connectivity to social networks \cite{lynn}. Jedlicka, in his review of the future of quantum neurobiology, touches on the evidence that quantum physics has also been used to describe human behaviour \cite{jedlicka, bruza,busemeyer}. The use of a quantum framework to accurately describe human behaviour does not necessarily mean that this behaviour results from quantum effects in neural processes. A clearer understanding of these processes, however, could also elucidate to what extent quantum physics is involved in the emergence of cognition from neural activity \cite{jedlicka}. Should it turn out that experimental evidence supports hypotheses of quantum enhanced neural processing we might turn this knowledge towards thinking about how we think.

\section*{Funding and Acknowledgments}
B.A and F.P. were supported by the South African Research Chair Initiative of the Department of Science and Technology and the National Research Foundation. B.A. was also supported by the National Institute for Theoretical Physics. Thank you to Angela Illing for the diagrams.
\bibliographystyle{aip}

\end{document}